\documentclass[deluxetables]{emulateapj}
\usepackage{apjfonts}
\usepackage{epsf}
\begin{document}
\slugcomment{}
\shortauthors{J. M. Miller et al.}
\shorttitle{The First Modern Stare at Mrk 817}

\title{The Inner Accretion Flow in the Resurgent Seyfert-1.2 AGN Mrk 817}

\author{J.~M.~Miller\altaffilmark{1},
A.~Zoghbi\altaffilmark{1},
M.~T.~Reynolds\altaffilmark{1},
J.~Raymond\altaffilmark{2},
D.~Barret\altaffilmark{3},
E.~Behar\altaffilmark{4},
W.~N.~Brandt\altaffilmark{5},
L.~Brenneman\altaffilmark{2},
P.~Draghis\altaffilmark{1},
E.~Kammoun\altaffilmark{1,3},
M.~J.~Koss\altaffilmark{6},
A.~Lohfink\altaffilmark{7},
D.~K.~Stern\altaffilmark{8}
}

\altaffiltext{1}{Department of Astronomy, University of Michigan, 1085
  South University Avenue, Ann Arbor, MI 48109-1107, USA,
  jonmm@umich.edu}
\altaffiltext{2}{Center for Astrophysics, Harvard \& Smithsonian, 60
  Garden Street, Cambridge, MA, 02138, USA}
\altaffiltext{3}{IRAP, Universit'{e} de Toulouse, CNRS, UPS, CNES 9,
  Avenue du Colonel Roche, BP 44346, F-31028, Toulouse Cedex 4,
  France}
\altaffiltext{4}{Department of Physics, Technion, Haifa, Israel}
\altaffiltext{5}{Department of Astronomy \& Astrophysics and the
  Institute for Gravitation and the Cosmos, The Pennsylvania State
  University, 525 Davey Lab, Univrersity Park, PA, 16802, USA}
\altaffiltext{6}{Eureka Scientific, 2452 Delmer Street Suite 100,
  Oakland, CA, 94602-3017, USA}
\altaffiltext{7}{Department of Physics, Montana State University,
  Bozeman, MT, 59717-3840, USA}
\altaffiltext{8}{Jet Propulsion Laboratory, California Institute of
  Technology, 4800 Oak Grove Drive, Pasadena, CA 91109, USA}

\begin{abstract}
Accretion disks and coronae around massive black holes have been
studied extensively, and they are known to be coupled.  Over a period
of 30 years, however, the X-ray (coronal) flux of Mrk 817 increased by
a factor of 40 while its UV (disk) flux remained relatively steady.
Recent high-cadence monitoring finds that the X-ray and UV continua in
Mrk 817 are also decoupled on time scales of weeks and months.  These
findings could require mechanical beaming of the innermost accretion
flow, and/or an absorber that shields the disk and/or broad line
region (BLR) from the X-ray corona.  Herein, we report on a 135~ks
observation of Mrk 817 obtained with NuSTAR, complemented by
simultaneous X-ray coverage via the Neil Gehrels Swift Observatory.
The X-ray data strongly prefer a standard relativistic disk reflection
model over plausible alternatives.  Comparable fits with related
models constrain the spin to lie in the range $0.5\leq a\leq 1$, and
the viewing angle to lie between $10^{\circ} \leq \theta \leq
22^{\circ}$ (including $1\sigma$ statistical errors and small
systematic errors related to differences between the models).  The
spectra also reveal strong evidence of moderately ionized absorption,
similar to but likely less extreme than obscuring events in NGC 5548
and NGC 3783.  Archival Swift data suggest that the absorption may be
variable.  Particularly if the column density of this absorber is
higher along the plane of the disk, it may intermittently mask or
prevent coupling between the central engine, disk, and BLR in Mrk 817.
\end{abstract}

\section{Introduction}
Extreme examples of variability in active galactic nuclei (AGN) are
increasingly common, thanks to ground-based monitoring and soft and
hard X-ray monitoring with Swift (e.g., Frederick et al.\ 2020,
Kaastra et al.\ 2014).  This poses some problems, but also some
opportunities.  A growing number of AGN are classified as
``changing-look'' AGN, for instance, and it is important to understand
whether these cases represent sudden geometric changes or evolution
that is consistent with state transitions in stellar-mass black holes
(e.g., Yang et al.\ 2018).  Separately, it is important but difficult
to distinguish flares due to regular accretion processes from genuine
tidal disruption events (TDEs; see, e.g., Trakhtenbrot et al.\ 2019,
van Velzen et al.\ 2021).  Sources that have undergone an extreme
evolution over relatively long time scales, and/or sources that may
still exhibit transitional geometrical phases, may offer key insights
into those that change rapidly.

Markarian 817 is a nearby Seyfert galaxy ($z=0.03145$; Strauss \&
Huchra 1988).  Recent work suggests that it should be classified as a
Seyfert 1.2 (e.g., Koss et al.\ 2017).  The active nucleus sits within
a barred spiral that is seen close to face-on, not atypical of
optically selected unobscured Seyferts (e.g., McLeod \& Rieke 1995,
Knapen et al.\ 2000).  Hubble images show evidence of dust along the
bar (Pogge \& Martini 2002).  The mass of the central black hole in
Mrk 817 has been determined using optical reverberation mapping:
$M_{BH} = 4.9\pm 0.8 \times 10^{7}~M_{\odot}$ (Peterson et al.\ 2004).

Winter et al.\ (2011) showed that the X-ray flux of Mrk 817 grew by a
factor of 40 between 1990 and 2011; however, the UV flux of the source
varied by only $\sim$ 2.3 over the same period.  This result is
difficult to explain in terms of standard disk and corona models,
since the disk (seen in UV) is expected to seed a Comptonizing corona
(seen in X-rays).  Recent monitoring with the Neil Gehrels Swift
Observatory confirms and extends this result: the X-ray and UV
continuum fluxes are also uncorrelated on time scales of just weeks
and days (Morales et al.\ 2019).

To better understand Mrk 817, it may be instructive to consider broad
absorption line quasars (BALQSOs).  The force multiplier effects that
enable radiation pressure on lines to expel gas at very high speeds
are only possible if the gas does not become too highly ionized (e.g.,
Proga 2003).  It now appears that this class is comprised of some
cases wherein the intrinsic X-ray flux is extremely low, and other
cases wherein the UV flow is shielded from the X-ray corona (Luo et
al.\ 2013, 2014; also see Baskin et al.\ 2014).  In the case of Mrk
817, the corona is no longer faint, but scattering or a shielding
geometry may be a plausible means of decoupling the UV and X-ray flux.

Relativistic reflection is often employed to measure black hole spin,
but it is also a powerful probe of the innermost accretion flow
geometry.  In Swift J1644$+$57, for instance, a broad but blue-shifted
Fe K line likely signals reflection from an outflowing funnel, defined
by a thick hyper-Eddington accretion disk (Kara et al.\ 2016).  In
less extreme cases, the emissivity profile can indicate
whether the corona is point-like or extended (see, e.g., Wilkins \&
Fabian 2012).  Alternatively, if a wind from the inner disk is
responsible for dampening feedback between the corona and disk, this
may be detected in X-ray absorption.

The NuSTAR observatory (Harrison et al. 2013) has proved to be an
extremely effective means of studying both relativistic reflection
(Miller et al.\ 2013a,b; Tomsick et al.\ 2014, Walton et al.\ 2014,
Zoghbi et al.\ 2015) and X-ray winds (e.g., King et al.\ 2014, Nardini
et al.\ 2015).  With the goal of understanding the innermost accretion
flow in Mrk 817, we obtained a deep observation of Mrk 817
with NuSTAR, complemented by a simultaneous Swift snapshot.  Owing to
the fact that Mrk 817 was much fainter than other Seyferts in the
formative years of X-ray astronomy, our NuSTAR observation represents
the first deep exposure of this source with a modern X-ray telescope.
The following section details the observations and how the data were
reduced.  The analysis and results are presented in Section 3.  We
discuss our results in Section 4.

\section{Observations and Data Reduction}
The long NuSTAR observation of Mrk 817 started on 18 December 2020, at
04:56:09 UTC (observation 60601007002).  The contemporaneous Swift
snapshot observation started on 19 December 2020, at 00:46:15
UTC (observation 0008904001).  The observations were reduced using the
tools in HEASOFT version 28, and the latest corresponding CALDB files
as of 4 January 2021.

NuSTAR data products were built from the cleaned FPMA and FPMB event
files, using the \texttt{nuproducts} routine.  Source events were
extracted from circular regions with radii of 120 arc seconds,
centered on the source position.  Background regions of equivalent
size were selected away from the source.  After filtering, the net
exposure time was 134.7~ks and the average count rate is 0.74~c/s
(FPMA and FPMB combined).  

Swift XRT products were built from the ``photon counting'' mode
data, owing to the modest flux from Mrk 817.  Source events were
extracted from a circular region with a radius of 30 pixels, centered
on the source; background events were extracted from an equivalent
region away from the source.  After filtering, the net exposure time
is 2.12~ks, and the average count rate is 0.095~c/s.

\section{Analysis \& Results}
The NuSTAR light curves of Mrk 817 show the variable ``flickering''
behavior that is typical of Seyferts.  There are no strong flares,
dips, or overall trends in the data; this is consistent with
expectations for a fairly massive black hole not undergoing
hyper-Eddington accretion (see Figure 1).  We therefore analyzed the
time-averaged spectra from the observations.  Low-energy differences
between the FPMA and FPMB (Madsen et al.\ 2020) have been resolved in
recent CALDB releases, so we combined the time-averaged FPMA and FPMB
spectra and responses using the tools \texttt{addascaspec} and
\texttt{addrmf}.

The spectra from NuSTAR and Swift have very different levels of
sensitivity, and require different binning methods in order to achieve
an optimal balance between continuum and line sensitivity.  After
experimenting with various schemes, we elected to bin the combined
NuSTAR spectrum to a minimum signal-to-noise ratio of 20 using the
tool \texttt{ftgrouppha} (based on the work of Kaastra \& Bleeker
2016).  The Swift/XRT spectrum only contains 200 counts, so we simply
grouped to require a minimum of 10 counts per bin (following Cash
1979) using the tool \texttt{grppha}.  Initial fits with
simple power-law models determined that the source is likely only
detected out to 50~keV; the NuSTAR spectrum was then fit
over the 3--50~keV band.  The Swift/XRT spectrum is insensitive above
7~keV, so it was fit over the 0.3--7.0~keV band.  The fits minimized
the $\chi^{2}$ statistic and the F-test was used to evaluate
improvements between nested models.

Figure 2 shows the spectra and data/model ratios obtained when the
data are jointly fit with a simple power-law model over the full band,
allowing the flux normalizations to vary.  The resulting fit is
unacceptable: $\chi^{2} = 710.7$ for $\nu = 154$ degrees of freedom
($\chi^{2}/\nu = 4.61$).  The ratios show strong evidence of a
relativistic Fe K emission line and disk reflection, as well as
evidence of low-energy absorption.  We therefore proceeded to model
the spectra in terms of relativistically blurred disk reflection,
modified by absorption that may be intrinsic to a disk wind or more
distant absorber.

To fit the reflection spectrum, we used \texttt{relxillD}, a variation
of the \texttt{relxill} model that allows the density of the disk to
vary (e.g., Garcia et al.\ 2014, Dauser et al.\ 2016).  The key
parameters of this model include the inner and outer emissivity
indices and the corresponding break radius (with a sign convention
given by $J \propto r^{-q}$), the dimensionless spin of the black hole
($a = cJ/GM^{2}$), the inclination at which the disk is viewed, the
inner and outer radii of the reflector, the systemic redshift, the
power-law index of the illuminating spectrum, the log of the
ionization parameter of the disk ($\xi = L/nr^{2}$), the abundance of
Fe relative to solar, the log of the number density of the disk, the
``reflection fraction'' (the relative importance of the reflection and
continuum), and the flux normalization of the component.  The
high-energy cut-off in this model is fixed at $E = 300$~keV.

Broadly following theoretical work (Wilkins \& Fabian 2012), we
constrained the inner emissivity to lie in the range $3\leq q_{in}
\leq 10$, the outer emissivity to lie in the range $0 \leq q_{out}
\leq 3$, and the break radius to lie within $2 \leq r_{br} \leq 6$ (in
units of $GM/c^{2}$).  The spin and inclination were allowed to vary
freely.  The inner disk radius was initially fixed at the innermost
stable circular orbit.  The outer disk radius was fixed at the maximum
value allowed by the model ($r_{out} = 990~GM/c^{2}$).  The power-law
index, ionization, Fe abundance, reflection fraction, and flux
normalization were all permitted to vary freely.  The reflector
density was initially fixed at ${\rm log n} = 15.0$, the same value
assumed in the default version of \texttt{relxill} and the related
\texttt{xillver} model.

To model the low-energy absorption, we employed \texttt{zxipcf}
(Reeves et al.\ 2008).  This is an absorption model based on
executions of XSTAR (e.g., Kallman et al.\ 2009), which is also the
basis of \texttt{relxill}.  The key parameters of this component are
the absorber column density ($N_{H}$), the log of the ionization
parameter ($\xi = L/nr^{2}$), and the velocity shift of the absorber.
The maximum allowed redshift was set to that of the host galaxy, but
was allowed to vary freely in the other direction (corresponding to
outflows in the frame of the host).

This model achieves an excellent fit to the spectrum, $\chi^{2}/\nu =
150.9/142 = 1.070$.  This is sufficiently close to unity that it
strongly implies that no additional model complexity is justified.
The measured X-ray flux in the 0.3--50~keV band is $F_{X} = 2.6(2)
\times 10^{-11}~ {\rm erg}~ {\rm cm}^{-2}~ {\rm s}^{-1}$ ($1\sigma$
errors are quoted throughout this work), implying an X-ray luminosity
of $L_{X} = 6.2(6) \times 10^{43}~{\rm erg}~ {\rm s}^{-1}$.  Assuming
a bolometric correction of $L_{bol}/L_{X} = 15$ (Vasudevan et
al.\ 2009), this translates to $L_{bol} \sim 9.2\times 10^{44}~{\rm
  erg}~ {\rm s}^{-1}$, or an Eddington fraction of $\lambda \sim
0.14$.

Table 1 lists the results of fits with this model.  The fit is shown
in Figure 3, alongside the underlying model.  The emissivity
parameters are fairly standard, and commensurate with expectations if
the corona is central and compact, and if the black hole has at least
a moderate spin (e.g., Wilkins \& Fabian 2012).  The spin parameter is
measured to be $a = 0.6^{+0.2}_{-0.1}$.  The $3\sigma$ upper limit on
the spin includes $a=0.99$, so maximal spin is not excluded.  However,
a spin of $a=0$ lies beyond the $4\sigma$ level of confidence.  The
inclination of the innermost disk is measured to be very low, $\theta
= 19^{+3}_{-4}$ degrees.  At $\Gamma = 2.10^{+0.07}_{-0.04}$, the
power-law index is within the range observed in samples of Seyferts
(Nandra et al.\ 2007).  The Fe abundance is found to be elevated, but
it is not strongly constrained, and this inference could be related to
the density assumed in the model (see, e.g., Jiang et al.\ 2019).  The
reflection fraction is also found to have a standard value, again
pointing to a typical innermost accretion flow geometry.

When the spectra are instead fit assuming a reflector density of ${\rm
  log}~n = 16.0$, the fit statistic is slightly worse ($\chi^{2} =
159.6$), but it becomes progressively worse at higher densities.  In
none of these cases does the best-fit iron abundance drop below
$A_{Fe} = 2$, however, suggesting that the fit is more sensitive to
the abundance than the density.  A conservative characterization of
our results is that the reflector density is in the ${\rm log}n \simeq
15-16$ range, and the Fe abundance is slightly but not dramatically
elevated relative to solar values.

When the \texttt{relxillD} model is replaced with \texttt{relxilllp},
which explicitly assumes a point-like emitter and lamp-post geometry,
with prescribed reflection fractions, an inferior fit is achieved
($\chi^{2}/\nu = 170.7/144$).  However, consistent values are measured
for the spin of the black hole and the inclination at which the inner
disk is viewed ($a=0.7^{+0.3}_{-0.2}$, $\theta=18\pm3$~deg.).  The
height of the lamp-post is measured to be $6\pm 1~GM/c^{2}$.  Unlike
the tests described above, this model is not nested, and it is not
clear that the F-statistic is appropriate.  However, using the F-test,
the \texttt{relxillD} model is only preferred at the 3.8$\sigma$ level
of confidence.  This could be regarded as tentative evidence that the
corona is not an idealized lamp-post, but still compact.  We note that
it is nominally possible to fit for the cut-off energy with
\texttt{relxilllp}, but the data are unable to constrain this
parameter well, so we fit with $E=300$~keV fixed.

To evaluate the level at which relativistic reflection is required
over distant reflection (from the BLR and/or torus), we replaced the
\texttt{relxill} with its basis model, \texttt{xillver}.  The
parameters of this model are the same as \texttt{relxill}, but it
lacks the spin and emissivity parameters that define relativistic
blurring.  Allowing all of the reflection and absorber parameters to
vary, a very poor fit is obtained: $\chi^{2}/\nu = 475.5/145$.  Again,
this model is not nested and the F-test is not necessarily
appropriate, but the difference in $\chi^{2}$ is enormous, indicating
that relativistic blurring is required at far more than the $8\sigma$
level of confidence.

The ionized absorber is strongly required by the spectra, but its
parameters are not well constrained; this is unsurprising given the
low resolution of the data.  Although the fit returns a very high
outflow velocity consistent with an ``ultra-fast outflow'' or UFO
($v/c\geq 0.1$; see, e.g., Tombesi et al.\ 2010), zero velocity shift
is within the 90\% confidence errors.  Moreover, it is very unlikely
that the resolution and sensitivity of the spectra could reliably
determine the velocity shift.  A conservative interpretation is that
the data allow for a UFO, but are consistent with a standard
low-velocity absorber.

The significance at which the absorber is required can be evaluated in
two ways.  First, dividing the best-fit column density by the
$1\sigma$ error suggests a significance of $8\sigma$.
Second, setting the column density to the minimum allowed by the model
($5\times 10^{20}~{\rm cm}^{-2}$) and re-fitting gives $\Delta\chi^{2}
= 161$ for $\Delta\nu = 1$, which is far more than the $8\sigma$ level
of confidence as determined by the F-test.

Significant low-energy absorption was not reported in the XMM-Newton
snapshot observation of Mrk 817 (Winter et al.\ 2011), nor in the
Swift monitoring exposures (Morales et al.\ 2019).  We therefore
examined various means by which absorption could be falsely implied:\\
\noindent$\bullet$ Cross calibration uncertainties between NuSTAR and
Swift could potentially skew the fitting results.  However, the
normalization of the \texttt{relxill} component in the Swift/XRT
spectrum is only 7\% lower than the NuSTAR spectrum in our model.
Moreover, if the normalizations are instead linked, absorption is even
more strongly required in the fit.

\noindent$\bullet$ Photon pile-up in CCD detectors leads to
artificially hard continua (Davis 2001).  To assess whether or not this
effect could have created a mismatch in the instrumental continua that
falsely implies absorption, we re-reduced the Swift/XRT data accepting
only ``grade=0'' events (single pixel strikes, typically free of
pile-up).  This spectrum shows no clear differences relative to
the standard spectrum.  It is therefore unlikely that spectral
hardening distortions due to pile-up have artificially created a break
in the spectrum that can be fit as absorption.

\noindent$\bullet$ It is possible that a flux deficit is artificially
created by failing to fit the continuum properly at low energy (note
that Winter et al.\ 2011 included a blackbody when modeling the
XMM-Newton snapshot observation).  We therefore added a disk blackbody
component to our model ($kT \leq 0.3$~keV).  A component of this kind
is not required by the data, either alone or in combination with the
absorber, and it does not serve to diminish the statistical
requirement for an absorber.

Instrumental and modeling issues do not appear to create a false
signature of absorption in the spectra, so we next examined whether or
not prior snapshot observations simply lacked the sensitivity needed
to detect significant absorption.  We used the Leicester Swift/XRT
spectra generator (Evans et al.\ 2009) to create a summed spectrum of
Mrk 817 from all observations taken throughout the mission, with a
total exposure of 213.4 ks.  Binning to require S/N = 10 and using the
same model, a good fit is again achieved when the absorber is included
($\chi^{2}/\nu = 334.9/309$).  The best-fit column density is measured
to be $N_{H} = 6.7(7)\times 10^{21}~{\rm cm}^{-2}$.  The covering
factor is measured to be $f = 0.44(9)$.  When the column density is
set to the model minimum (and/or when the covering factor is set to
zero), the fit statistic is markedly higher ($\Delta\chi^{2} = 107$).

The column and covering factor measured in the summed Swift/XRT
monitoring spectrum are {\it lower} than measured in the new deep
NuSTAR and Swift/XRT snapshot spectra (see Table 2).  This indicates
that the absorption is likely variable and diluted when measured in
the summed monitoring data.  These summed Swift/XRT spectrum also
permits no determination regarding a velocity shift; however, its
variable nature is consistent with the expected properties of a clumpy
outflow.

The absorber is most evident in the Swift/XRT band (see Figure 3), but
it has an effect even when the NuSTAR spectrum is considered alone.
When the absorber column density is set to its minimum and/or when the
covering factor is set to zero, the fit is rejected at the $7\sigma$
level of confidence ($\chi^{2}/\nu = 212.1/127$, relative to
$\chi^{2}/\nu = 139.4/126$ with absorption included).  The result is
equivalent when the FPMA and FPMB spectra are examined separately,
verifying that absorption is not implied owing to a calibration
mismatch between the FPMA and FPMB.

As noted by Reynolds et al.\ (2012), \texttt{zxipcf} samples the gas
ionization very coarsely (12 steps over nine orders of magnitude).
The data that we have examined lacks the sensitivity and resolution
needed to detect individual lines that would permit tight constraints
on the ionization parameter of the absorber; however, we checked our
results using the \texttt{warmabs} model.  This is a publicly
available, very high resolution model with tremendous flexibility,
also based on XSTAR (see, e.g., Kallman et al.\ 2014, Miller et
al.\ 2020).  To reproduce the partial covering parameter in
\texttt{zxipcf}, we constructed the model in the following manner:
$c_{1}\times warmabs\times relxillD + c_{2}\times relxillD$ (where the
relxillD parameters are fully linked, and $c_{1}$ and $c_{2}$ are
constants such that $c_{2} = 1.0-c_{1}$).

We find that broadly comparable fits ($\chi^{2} = 154/141$) to the
NuSTAR and Swift snapshot spectra are obtained using this model, and
that the absorption is required as strongly as before.  Similar values
are found for black hole spin ($a=0.7^{+0.3}_{-0.2}$), inclination ($i
= 15^{+3}_{-5}$~degrees), and emissivity, with overlapping confidence
intervals.  This model suggests that the column density could be
slightly lower ($N_{H} = 8\pm1 \times 10^{21}~ {\rm cm}^{-2}$) and
that the absorber may be more highly ionized (log$\xi =
1.5^{+0.2}_{-0.5}$); these values are not strongly excluded by the
prior model.  As before, this alternative model only weakly prefers a
high outflow velocity; it also prefers rms velocities as high as
$3000~{\rm km}~{\rm s}^{-1}$.  The fit statistic quoted here is for a
fit wherein the iron abundance was linked in the absorption and
reflection models, finding a value consistent with unity.  This
suggests that indications of enhanced abundances are likely not
robust.

As a final check, we constructed an X-ray hardness curve, using all
prior observations of Mrk 817 obtained by Swift.  We defined the soft
band as 0.3--2.0~keV, the hard band as 2.0--10.0~keV, and the hardness
as H$-$S/H$+$S.  We used the counts in these bands rather than the
flux, in order to avoid any dependence on models.  Neutral and
low-ionization absorption affects the low energy band, so intervals of
enhanced obscuration are marked by heightened hardness.  Figure 4
shows that Mrk 817 has recently displayed an unprecedented level of
spectral hardness, fully consistent with the enhanced obscuration
implied by our spectral fits.  These findings echo what may be
implicit evidence of strong, variable absorption based on changes in
the photon power-law index.  Winter et al.\ (2011) report changes of
$\Delta\Gamma \sim 0.6$ over a period of years; such variations are
more easily understood in terms of a standard corona if aided by
variable absorption.

We note that a search for X-ray reverberation from the inner disk
finds no clear signal.  At least in part, this is likely the result of
insufficient sensitivity: the fractional variability ($F_{var} \simeq
0.07$) and total counts obtained in the 3--10 keV band ($\sim4\times
10^{4}$) in our NuSTAR exposure are below the values typical of
observations wherein reverberation is detected (see, e.g., Figure 1 in
Kara et al.\ 2016).  However, if Mrk 817 truly lacks a relativistic
reflection signature in the Fe K band of its lag spectrum, it is not
unique.  The time-averaged spectra of other sources, such as
MCG-6-30-15, appear to require relativistic disk reflection but lack
the expected signal in the Fe~K band of their lag spectra (e.g. Kara
et al.\ 2016b).  It is possible that outflows from the innermost disk
serve to mute reverberation signatures in the same way that outflows
from larger radii wash out reverberation there.

\section{Discussion}
We have analyzed the first deep X-ray observation of Mrk 817 made with
a modern X-ray telescope.  The high sensitivity achieved with NuSTAR,
in combination with the extended pass band afforded by Swift, reveals
that the innermost accretion flow is consistent with a standard
compact hard X-ray corona and flat inner accretion disk.  This is
indicated by the fact that excellent fits to the broadband spectrum
are achieved using relativistic disk reflection.  Absorption is
strongly required in both the NuSTAR spectrum alone and in joint fits.
In examining the summed spectrum obtained over years of prior Swift
monitoring, there is also evidence of low-energy absorption, but at a
reduced level.  This likely indicates a variable outflow; this may be
the mechanism by which the UV and X-ray flux in Mrk 817 are sometimes
decoupled.  In this section, we examine these results in more detail
and place them into a broader context.

There is no indication of a funneled accretion flow or a corona with
unusual properties in Mrk 817.  The reflection spectrum is typical;
none of the measured model parameters are extreme.  A funneled,
outflowing inner geometry like that implied in Swift J1644$+$57 (e.g.,
Kara et al.\ 2016) and some weak emission-line quasars (e.g., Luo et
al.\ 2015) is ruled out.  This is unsurprising given that Mrk 817 is
sub-Eddington (our best fit model implies $\lambda = 0.14$ with a
standard bolometric correction) whereas Swift J1644$+$57 was likely in
a super- or hyper-Eddington phase when reflection was observed.  A
corona with emission that is primarily beamed away from the disk is
likely disfavored by the reflection fraction; a value well below unity
is expected if the corona is beamed away from the disk (Beloborodov
1999, Markoff \& Nowak 2005; Miller et al.\ 2012).  For these reasons,
it is unlikely that the structure of the disk and corona can account
for decoupled UV and X-ray fluxes in Mrk 817.

It is more likely that the absorption found in our spectral fits
interferes with disk--coronal coupling in Mrk 817.  Even in the
sources where reverberation from the optical BLR is best established,
there are periods wherein the link is broken.  In NGC 5548, for
instance, Kaastra et al.\ (2014) discovered and studied a transient
obscuration event (also see Dehghanian et al.\ 2019).  That
obscuration lasted for years, and was likely caused by a clumpy gas
stream that temporarily intersected our line of sight.  The novel
obscurer in NGC 5548 had column densities of $N_{H} \simeq 1\times
10^{22}~{\rm cm}^{-2}$ and low ionization parameters (${\rm log}\xi =
-1.2$, and ${\rm log}\xi \leq -2.1$, for each component).  Soon
thereafter, Mehdipour et al.\ (2017) reported a variable X-ray and UV
obscuring wind in NGC 3783, consistent with clumps at the base of the
optical BLR.  The two components in the obscuration partially cover
the source, have column densities of $N_{\rm H} = 0.3-2.0\times
10^{23}~{\rm cm}^{-2}$, and modest ionization parameters (${\rm
  log}\xi = 1.84$).  Unlike the case of NGC 5548, the obscuration in
NGC 3783 was observed to last only a month, suggestive of a passing
cloud in this source.

At least in terms of its column density and moderate ionization, the
low-energy X-ray absorption that we have tentatively discovered in Mrk
817 is similar to the variable obscuring events found in other AGN.
The fact that more modest obscuration is required in fits to the
summed spectrum from long-term Swift monitoring suggests that the
absorption is transient, which is again similar to NGC 5548 and NGC
3783.  It is reasonable to speculate, then, that the absorption in Mrk
817 is an outflow and potentially tied to the BLR.

This appears to be confirmed by a simple examination of the gas
properties.  Taking our estimate of the bolometric luminosity,
$L_{bol} \sim 9.2\times 10^{44}~{\rm erg}~ {\rm s}^{-1}$, and assuming
a plausible BLR cloud density of $1\times 10^{10}~{\rm cm}^{-3}$
(e.g., Gallo et al.\ 2021), the ionization parameter formalism ($\xi =
L/nr^{2}$) implies the obscuration in Mrk 817 is located at $r\simeq
1.7\times 10^{17}~{\rm cm}$, or $r\simeq 2.3\times 10^{4}~GM/c^{2}$.
This is fully consistent with the BLR.  The filling factor can then be
stimated via $N = n r f$ and our results imply $f\simeq 2.4\times
10^{-5}$.  Low volume filling factors are frequently inferred for the
BLR (e.g. Osterbrock 1991, Sneddon \& Gaskell 1999), and this is
fundamental to the success of reverberation mapping (e.g., Goad et
al.\ 2016).

Disk winds tend to be denser close to the plane of the disk, if only
because acceleration causes the gas density to fall.  We measured a
column density of $N_{H} = 4.0^{+0.7}_{-0.5}\times 10^{22}~{\rm
  cm}^{-2}$, for a reflection-derived inclination of $\theta =
19^{+3}_{-4}$~degrees.  An even higher density along the disk would be
even more effective at scattering or shielding interactions between
the corona and disk or BLR.  During intervals of low obscuration, it
may be easier to see the response of the BLR to the central engine,
facilitating reverberation measurements and black hole mass
constraints.  In other intervals, the variable obscuration may serve
to screen the disk and/or BLR from the central engine, leading to a
decouping of the UV and X-ray flux in Mrk 817.

Although our alternative model gives spin constraints that overlap
with our baseline model, it is best to combine the constraints on the
spin, and regard $0.5\leq a \leq 1$ as a confidence interval that
includes both statistical errors and small systematic errors between
related models.  If the spin of the black hole in Mrk 817 is indeed
moderate, it is consistent with emerging trends for a relatively high
black hole mass ($M_{BH} = 4.9\pm 0.8 \times 10^{7}~M_{\odot}$;
Peterson et al.\ 2004).  Observations tentatively indicate that
low-mass Seyferts may generally have high spin values, while
higher-mass Seyferts display a broader range of spins, perhaps
trending to lower values above $M_{BH} \geq 3\times 10^{7}~M_{\odot}$
(e.g., Reynolds 2021).  This tentative finding follows the expectation
that spin is more easily changed through accretion when the black hole
mass is lower, and that higher black hole masses may be partly
influenced by black hole mergers that can drive spins to lower values
(see, e.g., Berti \& Volonteri 2008, Fiacconi et al.\ 2018).

We have not detected a reverberation lag in Mrk 817 using NuSTAR.
However, a long program with XMM-Newton, or a coordinated program with
XMM-Newton and NuSTAR may succeed, particularly if a bright phase of
the state is targeted using a monitoring program.  In the future,
observations with XRISM (Tashiro et al.\ 2020) and Athena (e.g.,
Barret et al.\ 2018, Meidinger et al.\ 2020) may be able to study
reverberation lag spectra in Mrk 817 and a large number of similar
sources.

We acknowledge the anonymous referee for several helpful comments that
improved this paper.  We thank the {\it NuSTAR} director, Fiona
Harrison, and its lead mission planner, Karl Forster, for executing
this observation.  Similarly, we thank Brad Cenko and the {\it Swift}
team for making a supporting snapshot exposure.  EK acknowledges
financial support from the Centre National d'Etudes Spatiales (CNES).
JMM acknowledges helpful discussions with Jeremy Chen and Richard
Mushotzky.  This work made use of data from NuSTAR, a project led by
the California Institute of Technology, managed by the Jet Propulsion
Laboratory, and funded by NASA.

\clearpage


\begin{figure}
\hspace{1.0in}
\includegraphics[scale=0.7]{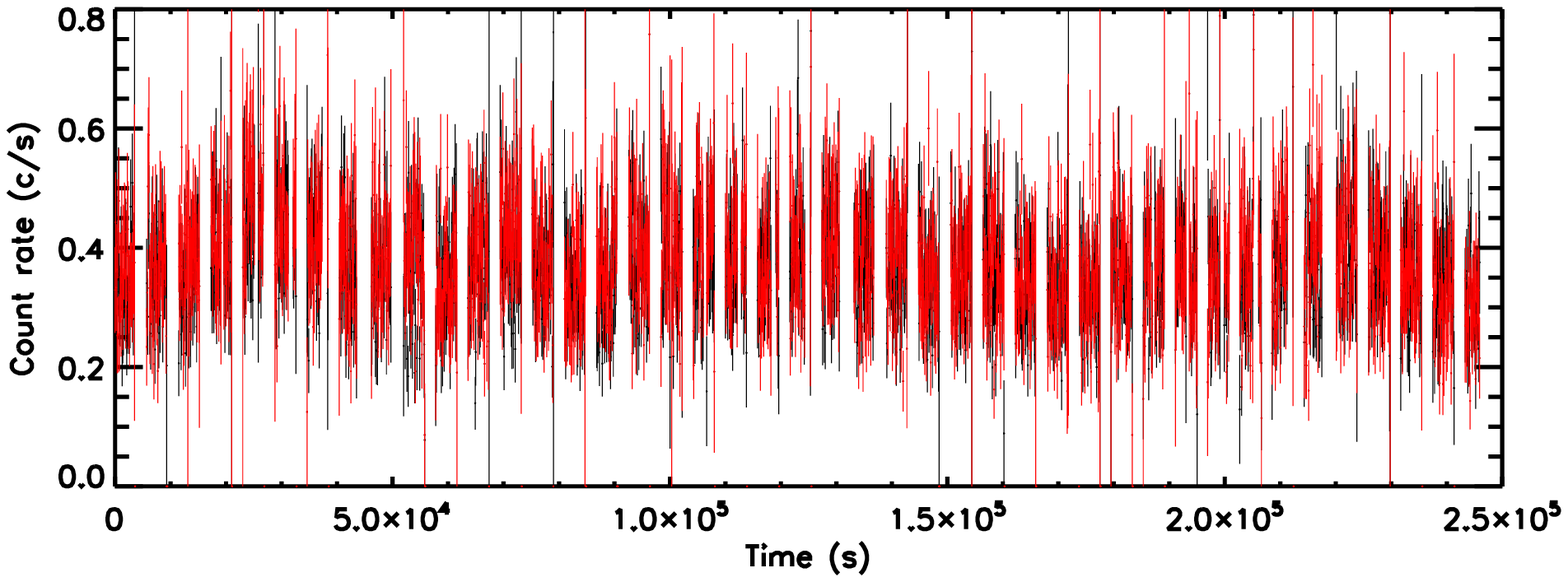}
\figcaption[t]{\footnotesize The NuSTAR FPMA (black) and FPMB (red)
  light curves of Mrk 817.  The full 3--79~keV count rate is plotted
  versus time; the time bins are 100~s.  Apparent flares or dips are
  simply edge effects from good time intervals.  The overall count
  rate shows only the ``flickering'' typical of accreting sources, and
  no clear trends.}
\end{figure}
\medskip

\clearpage

\begin{figure}
\hspace{1.0in}
\includegraphics[scale=0.5,angle=-90]{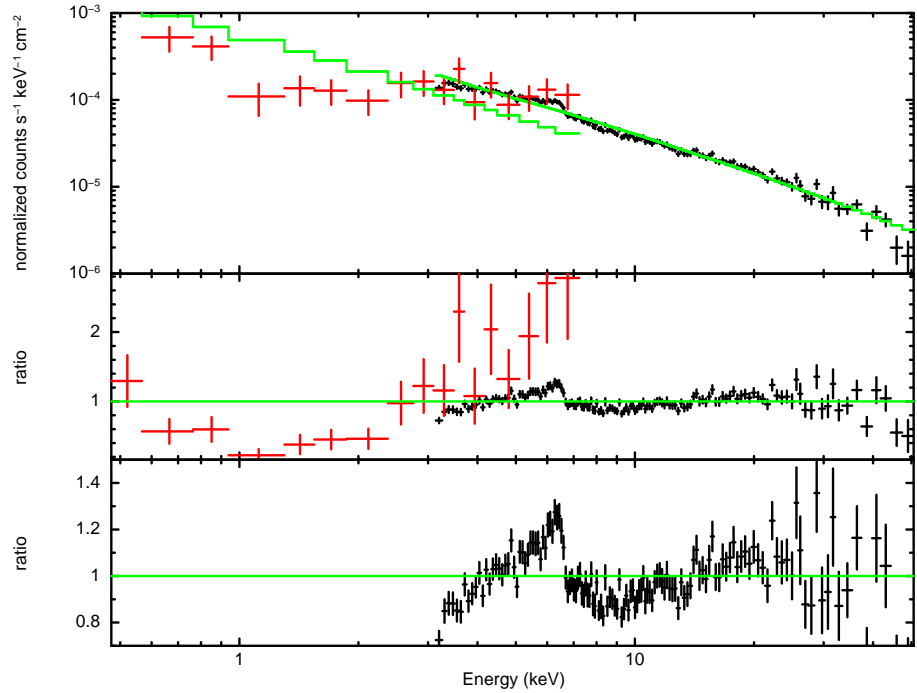}
\figcaption[t]{\footnotesize Top: the NuSTAR (black) and Swift (red)
  spectra of Mrk 817, jointly fit with a simple power-law model.
  Middle: the data/model ratios obtained with the power-law model.
  Bottom: the data/model ratio for the NuSTAR spectrum, on a scale
  that better illustrates clear signatures of relativistic disk
  reflection, including a skewed Fe~K emission line, and Compton
  back-scattering hump peaking in the 20--30~keV band.}
\end{figure}
\medskip

\begin{figure}
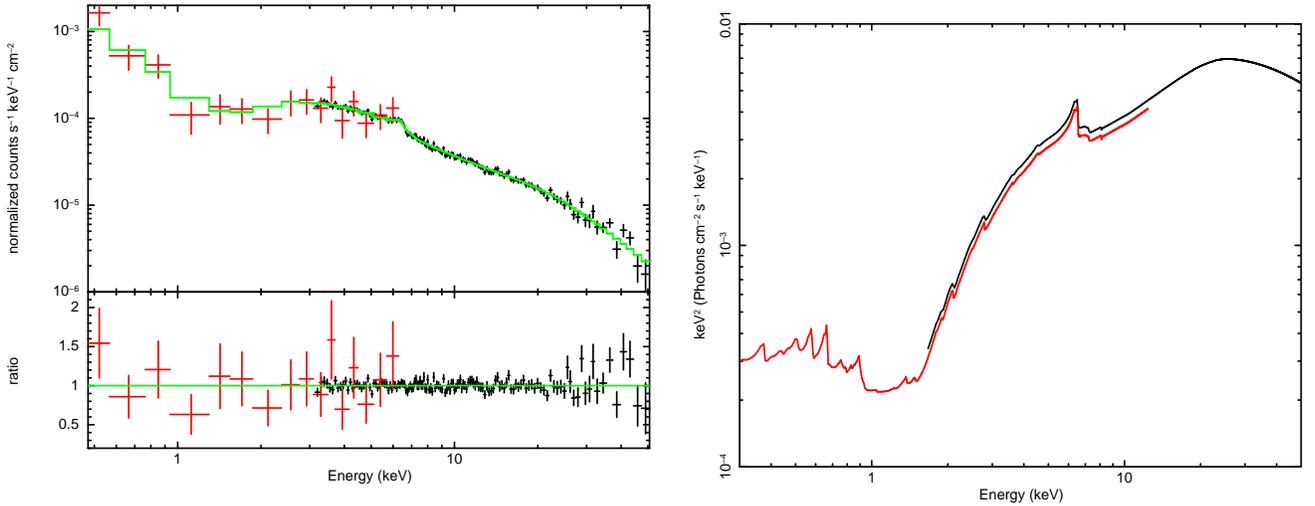

  \includegraphics[scale=0.35,angle=-90]{f3a.ps}
\includegraphics[scale=0.35,angle=-90]{f3b.ps}
\figcaption[t]{\footnotesize Left: The NuSTAR (black) and Swift (red)
  spectra of Mrk 817.  The spectra are fit with the model detailed in
  Table 1, and the corresponding data/model ratio is also shown.  The
  disk reflection spectrum is typical in all aspects, and the
  low-energy absorption may be tied to a transient outflow like those
  seen in sources like NGC 5548 and NGC 3783.  Right: The best-fit
  model for the NuSTAR (black) and Swift (red) spectra of Mrk 817,
  including low-energy absorption and relativistic disk reflection.
  The model is shown here without the data, to illustrate its features
  prior to convolution with the detector responses. }
\end{figure}
\medskip

\clearpage

\begin{figure}
\hspace{1.0in}
\includegraphics[scale=0.7]{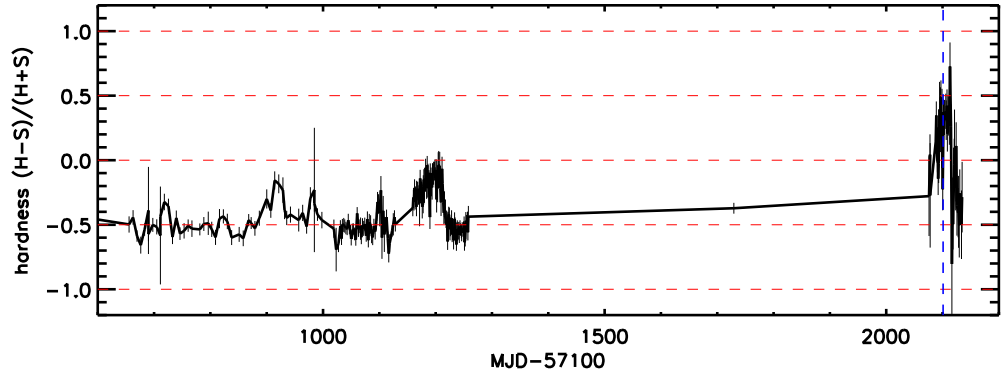}
\figcaption[t]{\footnotesize The Swift/XRT hardness curve of Mrk 817,
  covering the mission periods with dense monitoring.  The soft band
  is defined as 0.3--2.0~keV and the hard band is defined as
  2--10~keV.  The hardness curve is based on the number of counts in
  these bands, and is defined as (H$-$S)/(H$+$S).  The dashed blue
  line indicates the date of our NuSTAR exposure.  The unprecedented
  hardness at this time is consistent with obscuration acting in the
  soft X-ray band.}
\end{figure}
\medskip

\clearpage

\begin{table}[t]
\caption{Aborbed, blurred reflection parameters}
\begin{footnotesize}
\begin{center}
\begin{tabular}{lllllllllllll}
Component & log n & $q_{1}$ & $q_{2}$ & $r_{br}$ & $a$ & $\theta$ & $\Gamma$ & log$\xi$ & A$_{Fe}$ & $f_{refl}$ & $K_{refl}$ & $\chi^{2}/\nu$ \\
   ~  &   ~   &   ~     &   ~    & ($GM/c^{2}$) & ($cJ/GM^{2}$) & deg. & ~ & ~      &  ~      &    ~      & ($10^{-4}$) & ~ \\
\tableline
relxillD & 15.0* & 7(3) & 2.6(1) & $6_{-1}$ & $0.6^{+0.2}_{-0.1}$ & $19^{+3}_{-4}$ & $2.10^{+0.04}_{-0.07}$ & $1.3^{+0.4}_{-0.6}$ & $2.8^{+0.2}_{-0.5}$ & $1.3^{+0.3}_{-0.1}$ & $4.4^{+0.2}_{-0.1}$ & 150.1/141 \\
\tableline
~ & ~ & ~ & ~ & ~ & ~ & ~ & ~ & ~ & ~ & ~ & ~ & ~ \\
~ & $N_{H}$ & $f_{cov}$ & log$\xi$ & $v/c$ & ~ & ~ & ~ & ~ & ~ & ~ & ~ & ~ \\
  ~   & ($10^{22}~{\rm cm}^{-2}$) & ~ & ~ & ~ & ~ & ~ & ~ & ~ & ~ & ~ & ~ \\
\tableline
zxipcf & $4.0^{+0.7}_{-0.5}$ & $0.93(2)$ & $0.5^{+0.5}_{-0.4}$ & $-0.12^{-0.02}_{+0.09}$ & ~ & ~ & ~ & ~ & ~ & ~ & ~ \\
\tableline
\end{tabular}
\vspace*{\baselineskip}~\\
\end{center} 
\tablecomments{Parameters measured in joint fits to the Swift and
  NuSTAR spectra of Mrk 817 using a relativistically blurred disk
  reflection model (\texttt{relxillD}), modified by absorption
  (\texttt{zxipcf}).  The top row lists the parameters related to the
  reflection and continuum, as well as the fit statistic.  The
  emissivity parameters are listed first ($q_{1}$, $q_{2}$, and
  $r_{br}$), followed by the black hole spin and inner disk
  inclination, and finally the parameters tied to the continuum
  emission and reflector (the power-law index, $\Gamma$, and the disk
  ionization, iron abundance and reflection fraction.  The bottom row
  lists parameters related to the ionized absorber (the column
  density, covering factor, ionization parameter, and velocity in the
  frame of the host).  Parameters marked with an asterisk were frozed
  at the value listed.  Please see the text for additional details.}
\end{footnotesize}
\end{table}
\medskip


\end{document}